\documentclass[letterpaper,english,reprint, aps]{revtex4-1}

\usepackage[T1]{fontenc}
\usepackage[latin9]{inputenc}
\usepackage{booktabs}
\usepackage{textcomp}
\usepackage{amsmath}
\usepackage{amssymb}
\usepackage{graphicx}

\usepackage[colorlinks,
            linkcolor=blue,
            anchorcolor=blue,
            citecolor=blue,
            urlcolor=blue
            ]{hyperref}

\makeatletter

\pdfpageheight\paperheight
\pdfpagewidth\paperwidth

\makeatother

\usepackage{babel}

\begin{document}

\title{Topological Crystalline Insulators with $C_2$ Rotation Anomaly}

\author{Tan Zhang$^{1,2}$}
\author{Changming Yue$^{1,2}$}
\author{Tiantian Zhang$^{1,2}$}

\author{Simin Nie$^{3}$}
\altaffiliation {smnie@stanford.edu}
\author{Zhijun Wang$^{1,5}$}
\author{Chen Fang$^{1,2,4,5}$}
\altaffiliation {}
\author{Hongming Weng$^{1,2,4,5}$}
\altaffiliation {hmweng@iphy.ac.cn}
\author{Zhong Fang$^{1,2}$}

\affiliation{${^1}$Beijing National Research Center for Condensed Matter Physics, and Institute of Physics, Chinese Academy of Sciences, Beijing 100190, China}
\affiliation{${^2}$School of Physical Sciences, University of Chinese Academy of Sciences, Beijing 100049, China}
\affiliation{${^3}$Department of Materials Science and Engineering, Stanford University, Stanford, California 94305, USA}
\affiliation{${^4}$Songshan Lake Materials Laboratory , Dongguan, Guangdong 523808, China}
\affiliation{${^5}$CAS Center for Excellence in Topological Quantum Computation, University of Chinese Academy of Sciences, Beijing 100190, China}

\selectlanguage{english}
\begin{abstract}

Based on first-principles calculations and symmetry-based indicator analysis, we find a class of topological crystalline insulators (TCIs) with $C_2$ rotation anomaly in a family of Zintl
compounds, including $\mathrm{Ba}_{3}\mathrm{Cd}_{2}\mathrm{As}_{4}$, $\mathrm{Ba}_{3}\mathrm{Zn}_{2}\mathrm{As}_{4}$ and $\mathrm{Ba}_{3}\mathrm{Cd}_{2}\mathrm{Sb}_{4}$. The nontrivial
band topology protected by coexistence of  $C_2$ rotation symmetry and time-reversal symmetry $T$ leads to two surface Dirac cones at generic momenta on both top and bottom surfaces
perpendicular to the rotation axis. In addition, ($d-2$)-dimensional helical hinge states are also protected along the hinge formed by two side surfaces parallel with the rotation axis.
We develop a method based on nested Wilson loop technique to prove the existence of these surface Dirac cones due to $C_2$ anomaly and precisely locate them as demonstrated in studying these TCIs.
The helical hinge states are also calculated. Finally, we show that external strain can be used to tune topological phase transitions among TCIs, strong $Z_{2}$  topological insulators and
trivial insulators.
\end{abstract}
\maketitle

\emph{Introduction.}---The fermion multiplication theorem is a generalization of fermion doubling theorem in particle physics to crystalline solids. For example, in two-dimensional (2D) system with time-reversal symmetry of $T^2 =-1$ and $n$-fold rotational symmetry $C_{n}$ ($n$=2, 4, 6), this theorem imposes restriction on the number of stable massless Dirac fermions due to linear band crossing at arbitrarily generic momenta. The number of Dirac nodes must be a multiple of $2n$ in the first Brillouin zone (BZ) and they are robust against arbitrary perturbations preserving $T$ and $C_{n}$. Recently, quantum anomalies associated with discrete rotational symmetry $C_{n=2,4,6}$ and time-reversal symmetry $T$ have been proposed, which limit the number of massless fermions at generic momenta to be a multiple of $n$ instead of $2n$ in two dimension~\cite{2017arXiv170901929F}. Such rotation anomaly can only happen on the top or bottom surface of a three-dimensional (3D) system when the surfaces are perpendicular to the rotation axis~\cite{2017arXiv170901929F}. As we know, a single massless Dirac fermion with parity anomaly appears on the surface of a 3D topological insulator (TI) with time-reversal symmetry~\cite{PhysRevD.29.2375,PhysRevLett.53.2449,PhysRevLett.57.2967,PhysRevLett.61.2015,Fu2007}. The rotation anomaly leads to a new class of 3D TCIs~\cite{Fuannurev-conmatphys,RevModPhys.88.035005,PhysRevB.95.235425,2018arXiv181211959G}. Their surface states consist of $n$ Dirac cones, which evades the above fermion multiplication theorem. Furthermore, they support $n$ one-dimensional (1D) helical modes on the hinge formed by two side surfaces parallel with the rotation axis~\cite{PhysRevLett.119.246402,Benalcazar61,Schindlereaat0346,PhysRevLett.119.246401,PhysRevB.96.245115}.

There have been several materials proposed to have the above rotational anomaly. One of them is SnTe~\cite{PhysRevB.88.241303}, a well-known TCI protected by mirror symmetry with non-zero mirror Chern number. On its $(110)$ surface, the two surface Dirac cones, which have been thought to be protected by mirror symmetry, cannot be gapped if the mirror symmetry is broken while $C_2$ and $T$ are still preserved. Similarly, the four Dirac cones on the $(001)$ surface can also be understood to be protected by $C_4$ and $T$ symmetries~\cite{Hsieh2012}. The other example is anti-perovskite $\mathrm{Sr}_{3}\mathrm{Pb}\mathrm{O}$~\cite{PhysRevB.90.081112}. On its $(001)$ surface, there are four Dirac cones due to $C_4$ anomaly.
The $\mathrm{Ca}_{2}\mathrm{As}$ family are found to be TCIs protected by rotation symmetry, which have two Dirac cones at generic locations in momentum space if the mirror symmetry is broken, but the rotational and time-reversal symmetries are preserved~\cite{PhysRevB.98.241104}. 
Recently, one phase of BiBr compounds has also been proposed to be a TCI with $C_2$ rotation anomaly~\cite{Tang2019,Hsu_2019}, the surface Dirac cones and hinge states are also calculated. 

As a new class of TCIs, the corresponding topological invariant $\nu_{C_{n}}$ can be defined, although they are very difficult to be obtained. Recently, a convenient method
to identify these topologically nontrivial insulators has been proposed and programmed to screen all known nonmagnetic compounds~\cite{Song2018,Zhang2019}. The explicit and exhaustive mappings from symmetry data, band representations at high symmetric momenta, to topological indicators for arbitrary gapped band structure in the presence of time-reversal symmetry and all of the 230 space groups. The symmetry data of any gapped band structure can be compressed into a set of up to four $\mathbb{Z}_{m=2,3,4,6,8,12}$ numbers called symmetry-based indicators (SIs). These progresses~\cite{Zhang2019,Tang2019nature,Bernevig2019nature} have greatly accelerated the discovery of new topological materials.

\begin{figure}
\includegraphics{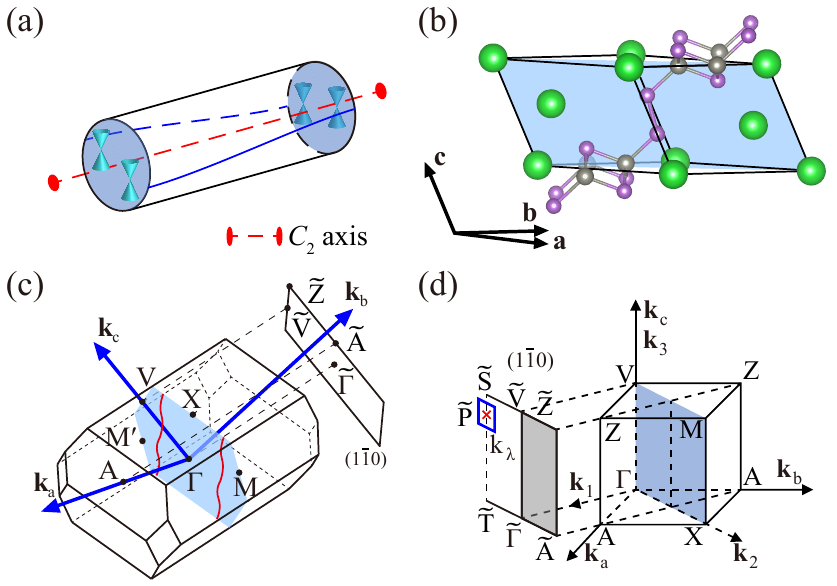}
\caption{(Color online) (a) The topological boundary states due to $C_{2}$ anomaly. For a cylinder like sample, there are two Dirac cones on both top and bottom surfaces and two 1D helical hinge states. (b) The primitive cell of $\mathrm{Ba}_{3}\mathrm{Cd}_{2}\mathrm{As}_{4}$. The green, grey and purple balls represent Ba, Cd and As atoms, respectively. The blue plane is the mirror plane. $\bold{a}$, $\bold{b}$ and $\bold{c}$ are three primitive lattice vectors.  (c) The first BZ of bulk and the $(1\bar{1}0)$ surface BZ. The mirror plane is marked as blue and the nodal lines in it are marked red. (d) The schematic diagram of BZ in reciprocal lattices $\bold{k}_a$, $\bold{k}_b$ and $\bold{k}_c$ and the redefined ones $\bold{k}_1$, $\bold{k}_2$ and $\bold{k}_3$, with $\bold{k}_1$ along $C_{2}$ rotation axis and $\bold{k}_2$, $\bold{k}_3$ being lattice vectors for (1$\bar{1}$0) surface BZ, where one of calculated surface Dirac cones is marked as red cross at $\widetilde{P}$ surrounded by a loop in blue lines.
\label{fig:1}}
\end{figure}

We find that three Zintl compounds, namely $\mathrm{Ba}_{3}\mathrm{Cd}_{2}\mathrm{As}_{4}$, $\mathrm{Ba}_{3}\mathrm{Zn}_{2}\mathrm{As}_{4}$ and $\mathrm{Ba}_{3}\mathrm{Cd}_{2}\mathrm{Sb}_{4}$, are TCIs classified by $C_2$ anomaly. We show the detailed and systematic method to judge a material with $C_{2}$ anomaly and prove the existing of surface Dirac cone with nested Wilson loop technique. They have two surface Dirac cones protected by $C_{2}$ and time-reversal symmetry $T$ at arbitrary momenta on both top and bottom $(1\bar{1}0)$ surfaces, as shown in Fig.~\ref{fig:1}. Besides, the two Dirac cones are connected by two helical hinge states on the side surface.
Compared with SnTe, $\mathrm{Sr}_{3}\mathrm{Pb}\mathrm{O}$ and $\mathrm{Ca}_{2}\mathrm{As}$ family, the proposed Zintl compounds have no mirror planes passing through $C_{2}$ rotation axis and the surface Dirac cones will be at generic momentum instead of being constrained by the mirror planes. This makes the $C_{2}$ rotation anomaly clearer. Comparing with BiBr, the difference is that the symmetry indicators is (0002) for BiBr, but (1102) for the Zintl compounds. The nontrivial $Z_2$ weak invariants indicate that the Zintl compounds have additional topological surface states.

\emph{Crystal structure and methodology.}---The Zintl compounds $\mathrm{Ba}_{3}\mathrm{Cd}_{2}\mathrm{As}_{4}$, $\mathrm{Ba}_{3}\mathrm{Zn}_{2}\mathrm{As}_{4}$ and $\mathrm{Ba}_{3}\mathrm{Cd}_{2}\mathrm{Sb}_{4}$ have been synthesized by Pb-flux methods and routine solid-state techniques in recent years~\cite{Wang2013,Saparov2008}, and possess abundant novel physical properties~\cite{Jiang2006,Holm2002,Fisher2000}. The crystallographic data and the atomic coordinates for these materials are listed in Supplemental Material \cite{SupplementalMaterial}.
They all crystallize in the same crystal structure and $\mathrm{Ba}_{3}\mathrm{Cd}_{2}\mathrm{As}_{4}$ is selected as an example to illustrate the structure, as shown in Fig.~\ref{fig:1}(b). The structure can be concisely described as stacked $\mathrm{Cd}_{2}\mathrm{As}_{4}$ layers which are separated by Ba cations.
The space group is $C2/m$ (No. 12) which has inversion symmetry $P$, rotation symmetry $C_2:(a,b,c)\rightarrow(-b,-a,-c)$ and mirror symmetry $M_{1\bar{1}0}:(a,b,c)\rightarrow(b,a,c)$.

For inversion symmetric system, the SIs $\mathbb{Z}_{2,2,2,4}=(z_{2,1},z_{2,2},z_{2,3},z_{4})$ are defined to indicate a new topological classification:
\begin{equation}
\label{eq:e1}
\begin{aligned}
z_{2,\varepsilon}\equiv&\sum_{\substack{\mathbf{K}\in\mathrm{TRIM} \\ \text{at } \{k_\varepsilon=\pi, \varepsilon=1,2,3\}}}\frac{n_-^\mathbf{K}-n_+^\mathbf{K}}{2}\;\mathrm{mod}\;2\\
z_{4}\equiv&\sum_{\mathbf{K}\in\mathrm{TRIM}}\frac{n_-^\mathbf{K}-n_+^\mathbf{K}}{2}\;\mathrm{mod}\;4
\end{aligned},
\end{equation}
where $n_\pm^\mathbf{K}$ is the number of occupied Kramer pairs having even (odd) parity at time reversal invariant momenta (TRIM). The three $z_{2,\varepsilon}$ invariants are equal to the three weak $Z_2$ topological invariants. When $z_{4}$ is 1 or 3, it indicates a strong TI, similar to the strong topological $Z_2$ invariant. When it is zero, the system is a weak TI if any one of $z_{2,\varepsilon}$ is nonzero, while $z_{4}$=2 means a TCI.

The method of calculating the Berry phase of the Wilson loop's eigen functions is used to prove the existence of surface Dirac cones at generic momenta. For a 3D system, its tight-binding Hamiltonian satisfies
\begin{equation}
\label{eq:e2}
\begin{aligned}
H(k_1,k_{\lambda})|u_{n}(k_1,k_{\lambda})\rangle=E_n(k_1,k_{\lambda}) |u_{n}(k_1,k_{\lambda})\rangle,
\end{aligned}
\end{equation}
where $k_1$ is along the periodical path in the bulk BZ perpendicular to the surface of loop $k_{\lambda}$.
We define a overlap matrix for $n_{occ}$ occupied states as $M_{mn}(k_{\alpha},k_{\alpha+1})=\langle u_{m}(k_{\alpha},k_{\lambda})|u_{n}(k_{\alpha+1},k_{\lambda})\rangle$ $(m,n=1,2\ldots n_{occ})$ ~\cite{PhysRevB.89.155114,Benalcazar61}.
Thus, the Wilson loop $\mathcal{W}(k_{\lambda})$ is defined as:
\begin{equation}
\label{eq:e3}
\begin{aligned}
\mathcal{W}(k_{\lambda})=\prod_{\alpha=0}^{N_1-1}M(k_{\alpha},k_{\alpha+1}),
\end{aligned}
\end{equation}
where the loop consisting of $N_1$ discretized $k_{\alpha}$ points with $k_{\alpha=N_1}=k_{\alpha=0}+G_1$ ($G_1$ is the length of the reciprocal lattice vector along $k_1$ loop). We can diagonalize $\mathcal{W}(k_{\lambda})$ and get the eigenvalue $W_{i}(k_{\lambda})=\exp(i\theta_{i}(k_{\lambda}))$ and the corresponding eigen vectors $\tilde{w}_{i}(k_{\lambda})$. $\theta_{i}(k_{\lambda})$ is referred as 1D hybrid Wannier center along $k_1$, which is also known as $i$-th Wilson loop flow along the loop $k_{\lambda}$~\cite{PhysRevB.47.1651}.

We define a new overlap matrix for selected $n^{\prime}$ number of Wilson eigen vectors as $\widetilde{M}_{\widetilde{m}\widetilde{n}}(k_{\lambda},k_{\lambda+1})=\langle \tilde{w}_{\widetilde{m}}(k_{\lambda})|\tilde{w}_{\widetilde{n}}(k_{\lambda+1})\rangle$ $(\widetilde{m},\widetilde{n}=1,2\ldots n^{\prime})$ ~\cite{Benalcazar61}, the nested Wilson loops along the closed loop consisting of $N^{\prime}$ discretized $k_{\lambda}$ points with $k_{\lambda=N^{\prime}}=k_{\lambda=0}$ are
\begin{equation}
\label{eq:e4}
\begin{aligned}
\widetilde{\mathcal{W}}=\prod_{\lambda=0}^{N^{\prime}-1}\widetilde{M}(k_{\lambda},k_{\lambda+1}).
\end{aligned}
\end{equation}
Further, Berry phase~\cite{Berry1984,vanderbilt2018berry} is written as
\begin{equation}
\label{eq:e5}
\begin{aligned}
\phi=-\operatorname{Im}\ln\det\widetilde{\mathcal{W}}.
\end{aligned}
\end{equation}
If the $n'$-th Wilson loop flow and the $(n'+1)$-th one cross each other and the crossing point is enclosed by the selected loop $k_{\lambda}$, the Berry phase $\phi$ from Eq.~\ref{eq:e5} will be $\pi$, or it will be zero.

\begin{figure}
\includegraphics{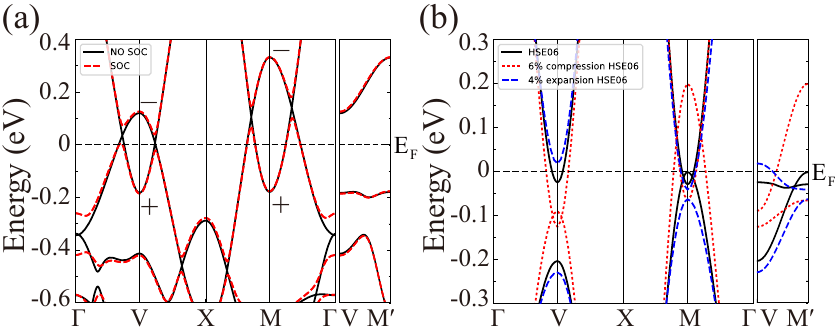}
\caption{(Color online) (a) The band structures of $\mathrm{Ba}_{3}\mathrm{Cd}_{2}\mathrm{As}_{4}$. The black solid and red dashed curves represent the bands from GGA calculations without and with SOC, respectively. The $+$ $(-)$ indicates the parity of two inverted bands at the $V$ and $M$ points.
(b) The band structures from HSE06 calculations. The black solid, red dotted and blue dashed curves represent the bands with no strain, 6\% compression and 4\% expansion of the lattice constants, respectively.
\label{fig:2}}
\end{figure}

\emph{Band structures and bulk topology.}---We choose $\mathrm{Ba}_{3}\mathrm{Cd}_{2}\mathrm{As}_{4}$ as an example in the following, since $\mathrm{Ba}_{3}\mathrm{Zn}_{2}\mathrm{As}_{4}$ and $\mathrm{Ba}_{3}\mathrm{Cd}_{2}\mathrm{Sb}_{4}$ have similar results as that of $\mathrm{Ba}_{3}\mathrm{Cd}_{2}\mathrm{As}_{4}$ and are presented in Supplemental Material \cite{SupplementalMaterial}. In the case without spin-orbit coupling (SOC) included, the calculated band structures within generalized gradient approximation
(GGA) show that two bands with opposite parity are inverted around the $V$ and $M$ points, as shown in Fig. \ref{fig:2}(a). The inverted band structures lead to nodal lines near Fermi level, which are protected by the time reversal symmetry $T$ and the inversion symmetry $P$~\cite{PhysRevB.84.235126,PhysRevB.92.045108,PhysRevB.95.045136}. Due to the constraint of the mirror symmetry $M_{1\bar{1}0}$, the crossing points form two 1D lines extending through the whole momentum space in the mirror plane.
When SOC is taken into account, all nodes along the nodal lines open band gaps, as shown by red dashed lines in Fig. \ref{fig:2}(a). Due to the band gaps induced by SOC, the topological invariants $Z_2$ are well defined by a \textquotedblleft curved chemical potential\textquotedblright{} which can be used to separate the valence and conduction bands. Its band topological invariants $Z_2=(0;110)$ can be easily calculated by the parity criterion proposed by Fu and Kane \cite{fu2007topological}, which indicates these materials are weak TIs. Besides, there are two surface Dirac cones on both $(001)$ and $(010)$ surfaces (see Supplemental Material  \cite{SupplementalMaterial} for more detail).

\emph{$C_{2}$ anomalous surface states.}---The SIs $\mathbb{Z}_{2,2,2,4}=(1102)$ for $\mathrm{Ba}_{3}\mathrm{Cd}_{2}\mathrm{As}_{4}$~\cite{Song2018,Zhang2019} have been obtained and two sets of topological invariants correspond to this SIs. Further calculation of the mirror Chern number $(\nu_{m_{1\bar{1}0}}=0)$ can distinguish them . We find the nontrivial topological invariant $\nu_{C_{2}} = 1$. The nonzero $\nu_{C_{2}}$ indicates that this material is a TCI with $C_2$ rotation anomaly, which hosts two surface Dirac cones locating on both top and bottom (1$\bar{1}$0) surfaces. If the sample is fabricated in a cylinder or prism along rotation axis [$1\bar{1}0$],  there will be two 1D helical hinge states related by $C_2$ rotation on the side surface as shown in Fig.~\ref{fig:1}(a).

\begin{figure}
\includegraphics{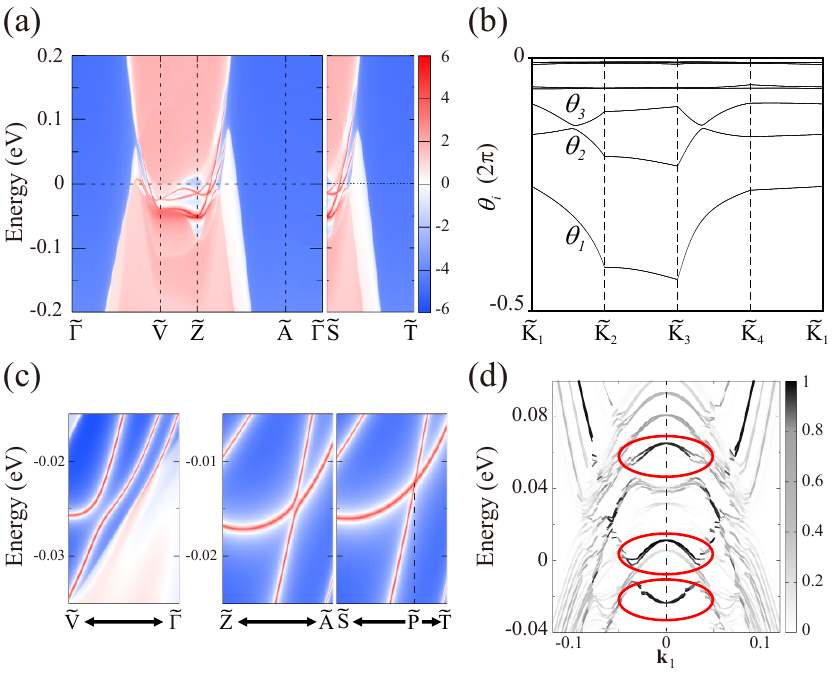}
\caption{(Color online) (a) The (1$\bar{1}$0) surface states of $\mathrm{Ba}_{3}\mathrm{Cd}_{2}\mathrm{As}_{4}$ obtained from the GGA+SOC calculations. (b) The Wilson loop flow along a chosen loop $\widetilde{K}_1$-$\widetilde{K}_2$-$\widetilde{K}_3$-$\widetilde{K}_4$-$\widetilde{K}_1$ which encloses a crossing point of two Wilson loop flows $\theta_2$ and $\theta_3$. (c) The enlarged plot of (1$\bar{1}$0) surface states along the $\widetilde{V}$-$\widetilde{\Gamma}$, $\widetilde{Z}$-$\widetilde{A}$ and $\widetilde{S}$-$\widetilde{T}$ paths. The surface Dirac cone is at $\widetilde{P}$ along $\widetilde{S}$-$\widetilde{T}$. (d) The helical hinge states are found along $\bold{k}_1$ direction. The darker color the more weight of the wave function on the two hinges where the hinge states exist. The darkest bands marked with red circles are hinge states.
\label{fig:3}}
\end{figure}

To well identify the Dirac cones on the (1$\bar{1}$0) surface, we calculate the surface states with modified onsite energy of atoms in the outmost unit cell by increasing 0.06 eV. This is reasonable and widely used to simulate the different chemical environment of atoms in and beneath the surface. The results are shown in Fig.~\ref{fig:3}(a). The surface states open gaps along both $\widetilde{\Gamma}$-$\widetilde{V}$ and $\widetilde{Z}$-$\widetilde{A}$ as shown in enlarged plots in Fig.~\ref{fig:3}(c). The Dirac cones at $\widetilde{V}$ and $\widetilde{Z}$ are trivial surface states since both of them can be pushed into valence or conduction bands without closing the band gap.
However, there are two surface Dirac cones at generic momenta due to $C_2$ anomaly on the $\widetilde{S}$-$\widetilde{T}$ path, which is off the high symmetrical line $\widetilde{Z}$-$\widetilde{A}$, as shown in Fig.~\ref{fig:3}(c). One surface Dirac cone locates at $\widetilde{P}$ (-0.4673,0.4375) on (1$\bar{1}$0) surface BZ and the other one is related by $C_{2}$ rotation.

To demonstrate the existence of surface Dirac cones on the (1$\bar{1}$0) surface, we calculate the Berry phase of the Wilson loop eigen functions based on the tight-binding Hamiltonian from the construction of maximally localized Wannier functions. Three new reciprocal lattice vectors $\bold{k}_1$, $\bold{k}_2$ and $\bold{k}_3$ as shown in Fig. \ref{fig:1}(d) have been defined, where $\bold{k}_1$ is along the $C_2$ rotation axis while $\bold{k}_2$ and $\bold{k}_3$ form the surface BZ. The loop integral of overlap matrix in Eq.~\ref{eq:e3} is along the reciprocal lattice $\bold{k}_1$ and the obtained Berry-Zak phase evolves along the loop $k_{\lambda}(k_2, k_3)$ in the (1$\bar{1}$0) surface BZ. The Wannier centers $\theta_{i}(k_{\lambda})$ and the eigen functions $\tilde{w}_{i}(k_{\lambda})$ are calculated with all occupied bands of $n_{occ}=22$ included in the effective Hamiltonian.
It is well-known that the Wilson loop flow have the same topological properties as the surface states. The Wilson loop flow along the high symmetrical paths in surface BZ are shown in Supplemental Material \cite{SupplementalMaterial}.
To identify the crossing point, we chose a $k_{\lambda}$ loop composed of $\widetilde{K}_1$, $\widetilde{K}_2$, $\widetilde{K}_3$, $\widetilde{K}_4$, which
in fact surrounds a crossing point of two Wilson loop flows $\theta_{2}$ and $\theta_{3}$ as shown in Fig. \ref{fig:3}(b). There is gap between $\theta_{2}$ and $\theta_{3}$ along the loop. To show there is a crossing point between $\theta_{2}$ and $\theta_{3}$ inside of the loop, two eigen functions $\tilde{w}_{i=1, 2}(k_{\lambda})$ are taken as occupied states to calculate the Berry phase along the loop according to Eq.~\ref{eq:e5}. Thus, we obtain $\pi$ Berry phase, which proves the existing of Dirac point.

\emph{Helical hinge states.}---As mentioned above, the helical 1D states can exist on the hinge where two gapped surfaces intersect each other. So that we take a prism geometry in Fig.~S5(a) in Supplemental Material \cite{SupplementalMaterial} with open boundary conditions around all four side surfaces and periodic boundary condition along the prism direction, namely $C_2$ rotation axis [1$\bar{1}$0].
We construct a tight-binding model of the new unit cell and calculate the band structure of a 6\texttimes{}6 supercell in Fig.~\ref{fig:3}(d). Because there is no direct band gap in bulk states, the band structure of supercell has no clear gap. To identify the hinge states, we calculate the weight of wave functions on the two hinges where the hinge states exist. The darkest bands are hinge states. The hinge states appear as a Dirac cone protected by $C_{2}$ and time-reversal symmetry at TRIM. They are localized at the two hinges (see the Supplemental Material \cite{SupplementalMaterial}). The 4\texttimes{}4 and 10\texttimes{}10 supercell are also calculated (see the Supplemental Material \cite{SupplementalMaterial}). We find the number of hinge sates does not change for  different sizes since the number of hinge is fixed, while the number of surface states and bulk states changes since the changes of surface area and bulk volume for different sizes. The hinge states in the other prism geometry are also calculated, and the results are similar as shown in the Supplemental Material \cite{SupplementalMaterial}. The hinge states are buried within the bulk and surface states. However, since hinge states have dominant local density of states along hinges, they might be measured by STM as similar as the hinge states measurement of bismuth \cite{Schindler_2018}.

\begin{figure}
\includegraphics[clip,scale=1, angle=0]{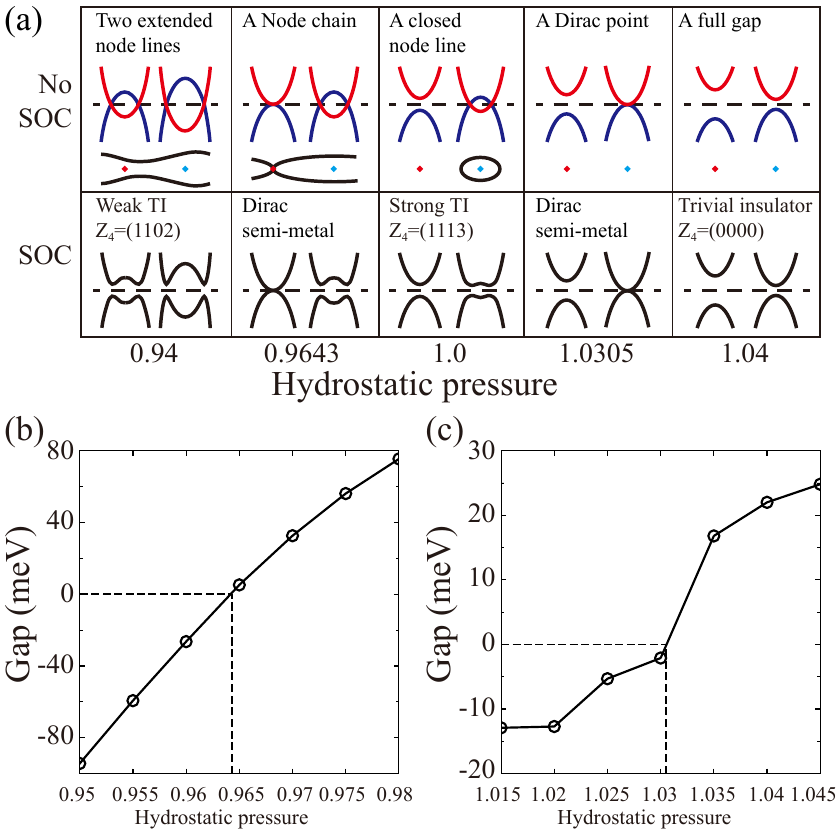}
\caption{(Color online) (a) The topological phase transitions of $\mathrm{Ba}_{3}\mathrm{Cd}_{2}\mathrm{As}_{4}$ under different hydrostatic pressure. The pressure is simulated by changing three lattice constants in the ratio of $a_{0}$ and $c_{0}$ within HSE06 calculations. The upper row shows bands (red and blue curves) close to the Fermi energy (black dashed lines) without SOC at $V$ (red point) and $M$ (cyan point). The black curves indicate the nodal lines around them. The lower row shows bands (black curves) with SOC included. (b) and (c) are the band gap change without SOC at $V$ point under compression and at $M$ under expansion, respectively. The dashed lines indicate the phase transition critical points where the band gap closes.
\label{fig:4} }
\end{figure}

\emph{Phase diagrams and topological phase transitions.}---Considering the well-known underestimation of band gap within GGA, additional hybrid functional calculations in  Heyd-Scuseria-Ernzerhof (HSE06) scheme are performed to check the band inversion in $\mathrm{Ba}_{3}\mathrm{Cd}_{2}\mathrm{As}_{4}$. As shown in Fig.~\ref{fig:2}(b), the band inversion within HSE06 only happens at $M$ instead of both $V$ and $M$ within GGA. In view of the small band inversion depth, the band inversion might be adjusted by external strain.
When the lattice constants are compressed by 6\%, the band inversion appears at both $V$ and $M$. However, when the lattice constants are expanded by 4\%, there is no band inversion at either $V$ or $M$. We also get similar results for $\mathrm{Ba}_{3}\mathrm{Zn}_{2}\mathrm{As}_{4}$ and $\mathrm{Ba}_{3}\mathrm{Cd}_{2}\mathrm{Sb}_{4}$ in the Supplemental Material \cite{SupplementalMaterial}.

Therefore, we can tune these materials into various topologically distinct states by compression or expansion of lattice constants in the way of hydrostatic pressure.
The phase diagrams about the bands at $V$ and $M$, and the nodal lines in $\mathrm{Ba}_{3}\mathrm{Cd}_{2}\mathrm{As}_{4}$ are shown in Fig.~\ref{fig:4}(a) under different strain. In the case without SOC, the band inversion only happens at $M$ under zero pressure with lattice constants being $a_{0}$, $b_{0}$ and $c_{0}$.
When the lattice constants are compressed, the band inversion increases at $M$ and the band gap closes at $V$, and band inversion appears at $V$ as shown in Fig.~\ref{fig:4}(b). The critical point of the phase transition is at $a=b=0.9643a_{0}$, $c=0.9643c_{0}$. When the lattice constants are expanded, the band gap increases at $V$, while the band inversion disappears at $M$, as shown in Fig. \ref{fig:4}(c). The critical point is at $a=b=1.0305a_{0}$, $c=1.0305c_{0}$.
When SOC is considered, the $\mathrm{Ba}_{3}\mathrm{Cd}_{2}\mathrm{As}_{4}$ is a strong TI with $\mathbb{Z}_{2,2,2,4}=(1113)$ without pressure. It transforms from a strong TI to a TCI with $\mathbb{Z}_{2,2,2,4}=(1102)$, and to a normal insulator, respectively, and it becomes a Dirac semimetal at two critical points of the phase transition. The similar results can also be obtained for $\mathrm{Ba}_{3}\mathrm{Zn}_{2}\mathrm{As}_{4}$ and $\mathrm{Ba}_{3}\mathrm{Cd}_{2}\mathrm{Sb}_{4}$.


\emph{Conclusion.}---We demonstrate a new class of TCI with $C_2$ rotation anomaly in Zintl compounds ($\mathrm{Ba}_{3}\mathrm{Cd}_{2}\mathrm{As}_{4}$, $\mathrm{Ba}_{3}\mathrm{Zn}_{2}\mathrm{As}_{4}$ and $\mathrm{Ba}_{3}\mathrm{Cd}_{2}\mathrm{Sb}_{4}$) by first-principles calculations and SI analysis.
When SOC is ignored, these materials are nodal line semimetals with two extended nodal lines in the mirror plane within GGA calculation.
With consideration of SOC, they become TCIs. There are only two surface Dirac cones protected by $C_{2}$ and time-reversal symmetry $T$ at arbitrary momenta on both top and bottom $(1\bar{1}0)$ surfaces. The precise positions of surface Dirac cones are determined by calculations of $(1\bar{1}0)$ surface states. We develop a nested Wilson loop method to prove the existence of surface Dirac cones. The helical hinge states on the side surface are also calculated. Within HSE06 calculations, we get the topological phase diagrams of $\mathrm{Ba}_{3}\mathrm{Cd}_{2}\mathrm{As}_{4}$ under compression and expansion hydrostatic pressure. This new class of TCIs are experimentally synthesized and to be verified by experiments. It provides us an ideal platform to study the $C_2$ rotation anomaly and high order TIs with hinge states.

We acknowledge the supports from the Ministry of Science and Technology of China (Grants No. 2016YFA0300600, 2016YFA0302400 and 2018YFA0305700), the National Natural Science Foundation (Grant No. 11674370, 11421092 and 11674369), the Chinese Academy of Sciences (Grant No. XDB28000000 and XXH13506-202), the Science Challenge Project (TZ2016004), the K. C. Wong Education Foundation (GJTD-2018-01), the Beijing Natural Science Foundation (Z180008), and the Beijing Municipal Science and Technology Commission (Z181100004218001).

\bibliography{paperref}

\end{document}


\title{Supplemental Material \--- Topological Crystalline Insulators with $C_2$ Rotation Anomaly}

\author{Tan Zhang$^{1,2}$}
\author{Changming Yue$^{1,2}$}
\author{Tiantian Zhang$^{1,2}$}

\author{Simin Nie$^{3}$}
\altaffiliation {smnie@stanford.edu}
\author{Zhijun Wang$^{1,5}$}
\author{Chen Fang$^{1,2,4,5}$}
\altaffiliation {}
\author{Hongming Weng$^{1,2,4,5}$}
\altaffiliation {hmweng@iphy.ac.cn}
\author{Zhong Fang$^{1,2}$}

\affiliation{${^1}$Beijing National Research Center for Condensed Matter Physics, and Institute of Physics, Chinese Academy of Sciences, Beijing 100190, China}
\affiliation{${^2}$School of Physical Sciences, University of Chinese Academy of Sciences, Beijing 100049, China}
\affiliation{${^3}$Department of Materials Science and Engineering, Stanford University, Stanford, California 94305, USA}
\affiliation{${^4}$Songshan Lake Materials Laboratory , Dongguan, Guangdong 523808, China}
\affiliation{${^5}$CAS Center for Excellence in Topological Quantum Computation, University of Chinese Academy of Sciences, Beijing 100190, China}

\selectlanguage{english}

\maketitle

\subsection{Crystal structure and methodology}
$\mathrm{Ba}_{3}\mathrm{Cd}_{2}\mathrm{As}_{4}$,
$\mathrm{Ba}_{3}\mathrm{Zn}_{2}\mathrm{As}_{4}$ and $\mathrm{Ba}_{3}\mathrm{Cd}_{2}\mathrm{Sb}_{4}$ have the same crystal structure.
We choose $\mathrm{Ba}_{3}\mathrm{Cd}_{2}\mathrm{As}_{4}$ as an example and show its structure in Fig  \ref{spfig:1}.
The crystallographic data and the atomic coordinates for these three materials are listed in Tables \ref{tab:1} and \ref{tab:2}, respectively.

The first-principles calculations are carried out with the Vienna ab initio simulation package (VASP) \cite{kresse1996efficiency,kresse1996efficient}. The exchange-correlation potential is treated using the generalized gradient approximation (GGA) parametrized by Perdew, Burke, and Ernzerhof (PBE) \cite{Perdew1996}. The cutoff energy is set to 450 eV, and the Brillouin zone (BZ) is sampled with 10 \texttimes{} 10 \texttimes{} 10 Monkhorst-Pack k-point mesh. The Heyd-Scuseria-Ernzerhof (HSE06) calculations \cite{Heyd2003} are further applied to check the band gap and band inversion. In order to explore the surface states of the Zintl compounds, maximally localized Wannier functions for the $s$ orbits of Cd (Zn) atoms and the $p$ orbits of As (Sb) atoms have been constructed and used to build the tight-binding model \cite{Mostofi2008,Marzari2012}. The projected surface states are obtained from the tight-binding Hamiltonian by Green function method \cite{WU2017}.

\setcounter{figure}{0}
\makeatletter
\renewcommand{\thefigure}{S\@arabic\c@figure}
\makeatother

\setcounter{table}{0}
\makeatletter
\renewcommand{\thetable}{S\@arabic\c@table}
\makeatother

\begin{figure}
\includegraphics{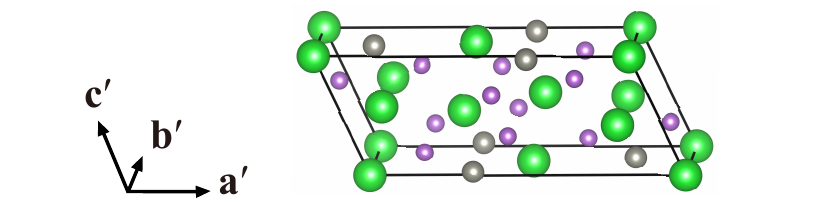}
\caption{(Color online) The conventional unit cell of $\mathrm{Ba}_{3}\mathrm{Cd}_{2}\mathrm{As}_{4}$. The green, grey and purple balls represent Ba, Cd and As atoms, respectively. $\bf{a'}$ , $\bf{b'}$ and $\bf{c'}$ are three conventional lattice vectors.
\label{spfig:1}}
\end{figure}

\begin{table}
\caption{Crystallographic data for $\mathrm{Ba}_{3}\mathrm{Cd}_{2}\mathrm{As}_{4}$,
$\mathrm{Ba}_{3}\mathrm{Zn}_{2}\mathrm{As}_{4}$ and $\mathrm{Ba}_{3}\mathrm{Cd}_{2}\mathrm{Sb}_{4}$.
\label{tab:1}}

\begin{tabular*}{1\columnwidth}{@{\extracolsep{\fill}}cccc}
\toprule
Formula & $\mathrm{Ba}_{3}\mathrm{Cd}_{2}\mathrm{As}_{4}$\cite{Wang2013} & $\mathrm{Ba}_{3}\mathrm{Zn}_{2}\mathrm{As}_{4}$\cite{Wang2013} & $\mathrm{Ba}_{3}\mathrm{Cd}_{2}\mathrm{Sb}_{4}$\cite{Saparov2008} \tabularnewline
\midrule
$a^{\prime}$ ($\mathring{A}$) & 17.089(3) & 16.916(4) & 17.835(2)\tabularnewline
$b^{\prime}$ ($\mathring{A}$) & 4.6076(7) & 4.497 (1) & 4.8675(5)\tabularnewline
$c^{\prime}$ ($\mathring{A}$) & 7.304(1) & 7.225(2)  & 7.6837(7)\tabularnewline
$\alpha$ (deg) & 90 & 90 & 90\tabularnewline
$\beta$ (deg) & 112.312(1) & 113.147(2) & 112.214(1)\tabularnewline
$\gamma$ (deg) & 90 & 90 & 90\tabularnewline
\bottomrule
\end{tabular*}
\end{table}

\begin{table}
\caption{Atomic coordinates and equivalent isotropic displacement parameters
for $\mathrm{Ba}_{3}\mathrm{Cd}_{2}\mathrm{As}_{4}$, $\mathrm{Ba}_{3}\mathrm{Zn}_{2}\mathrm{As}_{4}$ and $\mathrm{Ba}_{3}\mathrm{Cd}_{2}\mathrm{Sb}_{4}$.\label{tab:2}}

\begin{tabular*}{1\columnwidth}{@{\extracolsep{\fill}}cccccc}
\toprule
Atoms &  & Wyckoff & $x$ & $y$ & $z$\tabularnewline
\midrule
$\mathrm{Ba}_{3}\mathrm{Cd}_{2}\mathrm{As}_{4}$ & Ba1 & 4i & 0.14204(3) & 0 & 0.57734(7)\tabularnewline
 & Ba2 & 2a & 0 & 0 & 0\tabularnewline
 & Cd1 & 4i & 0.33048(4) & 0 & 0.02867(9)\tabularnewline
 & As1 & 4i & 0.20742(5) & 0 & 0.1918(1)\tabularnewline
 & As2 & 4i & 0.49436(5) & 0 & 0.3254(1)\tabularnewline
 $\mathrm{Ba}_{3}\mathrm{Zn}_{2}\mathrm{As}_{4}$ & Ba1 & 4i & 0.14370(3) & 0 & 0.58144(7)\tabularnewline
 & Ba2 & 2a & 0 & 0 & 0\tabularnewline
 & Zn1 & 4i & 0.32838(6) & 0 & 0.0366(1)\tabularnewline
 & As1 & 4i & 0.20638(5) & 0 & 0.1673(1)\tabularnewline
 & As2 & 4i & 0.48895(5) & 0 & 0.3203(1)\tabularnewline
$\mathrm{Ba}_{3}\mathrm{Cd}_{2}\mathrm{Sb}_{4}$ & Ba1 & 4i & 0.14194(2) & 0 & 0.59031(5)\tabularnewline
 & Ba2 & 2a & 0 & 0 & 0\tabularnewline
 & Cd1 & 4i & 0.33167(3) & 0 & 0.02826(6)\tabularnewline
 & Sb1 & 4i & 0.20771(2) & 0 & 0.19654(6)\tabularnewline
 & Sb2 & 4i & 0.49775(2) & 0 & 0.31514(5)\tabularnewline
\bottomrule
\end{tabular*}
\end{table}

\subsection{The band structures of $\mathrm{Ba}_{3}\mathrm{Zn}_{2}\mathrm{As}_{4}$ and $\mathrm{Ba}_{3}\mathrm{Cd}_{2}\mathrm{Sb}_{4}$}

\begin{figure}
\includegraphics{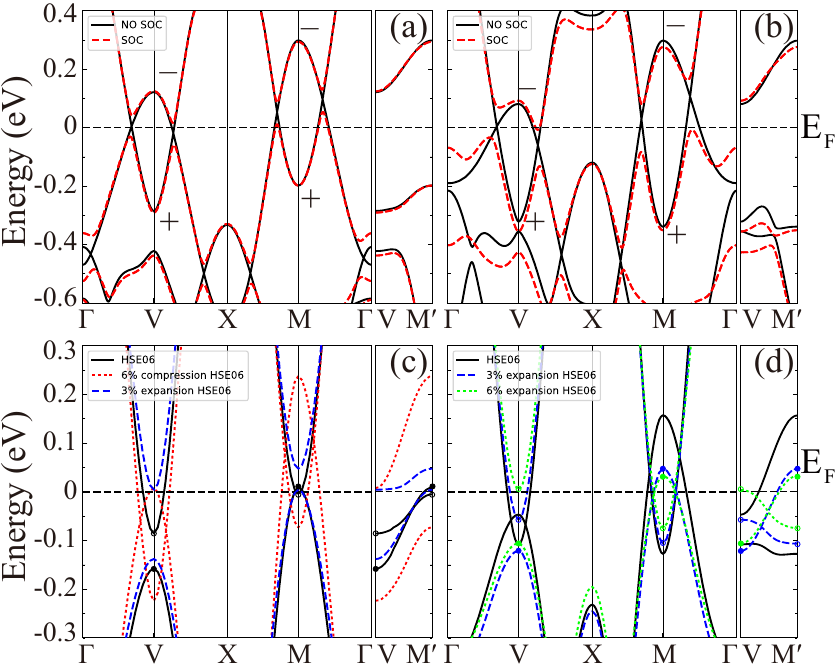}
\caption{(Color online) The band structures of (a) $\mathrm{Ba}_{3}\mathrm{Zn}_{2}\mathrm{As}_{4}$ and (b) $\mathrm{Ba}_{3}\mathrm{Cd}_{2}\mathrm{Sb}_{4}$ by GGA calculations. The black solid and red dashed curves represent the bands obtained without and with SOC, respectively. The $+$ $(-)$ indicates the parity of two inverted bands at the $V$ and $M$ points.
The HSE06 band structures of (c) $\mathrm{Ba}_{3}\mathrm{Zn}_{2}\mathrm{As}_{4}$ and (d) $\mathrm{Ba}_{3}\mathrm{Cd}_{2}\mathrm{Sb}_{4}$ without SOC. The black solid, red dotted, blue dashed and green dotted curves represent the bands of the materials with no strain, 6\% compression, 3\% expansion and 6\% expansion of lattice constants, respectively. The dots and circles mark the different bands where the band inversion emerges.
\label{spfig:2}}
\end{figure}

Figs. \ref{spfig:2}(a) and (b) are the band structures of $\mathrm{Ba}_{3}\mathrm{Zn}_{2}\mathrm{As}_{4}$ and $\mathrm{Ba}_{3}\mathrm{Cd}_{2}\mathrm{Sb}_{4}$ with and without spin-orbit coupling (SOC) within GGA calculation, respectively, which are similar to the results of $\mathrm{Ba}_{3}\mathrm{Cd}_{2}\mathrm{As}_{4}$.

Considering the well-known underestimation of band gap within GGA, the HSE06 calculation without SOC are performed to check the band inversion at the $V$ and $M$ points, as shown as the black solid lines in Figs. \ref{spfig:2}(c) and (d). The band inversion exists only at $M$ point for $\mathrm{Ba}_{3}\mathrm{Zn}_{2}\mathrm{As}_{4}$, while it still appears at both $V$ and $M$ points for $\mathrm{Ba}_{3}\mathrm{Cd}_{2}\mathrm{Sb}_{4}$.
The band inversion  under pressure has been studied too. For $\mathrm{Ba}_{3}\mathrm{Zn}_{2}\mathrm{As}_{4}$, when the lattice constants are compressed by 6\%, the band inversion appears at $V$ point,  in addition the band inversion at $M$ point increases. However, when the lattice constants are expanded by 3\%, the band inversion at $M$ point is removed and it becomes a trivial insulator, as shown in Fig. \ref{spfig:2}(c).
For $\mathrm{Ba}_{3}\mathrm{Cd}_{2}\mathrm{Sb}_{4}$, 3\% or 6\% expansion of the lattice constants will remove the band inversion at $V$ point, but the band inversion at $M$ point still exists, as shown in Fig. \ref{spfig:2}(d).

\subsection{The $(001)$ and $(010)$ surface states of $\mathrm{Ba}_{3}\mathrm{Cd}_{2}\mathrm{As}_{4}$, $\mathrm{Ba}_{3}\mathrm{Zn}_{2}\mathrm{As}_{4}$ and $\mathrm{Ba}_{3}\mathrm{Cd}_{2}\mathrm{Sb}_{4}$}

\begin{table}
\caption{The parity productions $\delta_{i}$ at eight TRIM points. The bands are obtained from GGA calculation of experimental crystal structure with SOC included.\label{tab:3}}

\begin{tabular*}{1\columnwidth}{@{\extracolsep{\fill}}ccc}
\toprule
$\Gamma(0,0,0)$ & $2A(\frac{1}{2},0,0)$ & $X(\frac{1}{2},\frac{1}{2},0)$\tabularnewline
$-$ & $+$ & $-$\tabularnewline
\midrule
$V(0,0,\frac{1}{2})$ & $2Z(\frac{1}{2},0,\frac{1}{2})$ & $M(\frac{1}{2},\frac{1}{2},\frac{1}{2})$\tabularnewline
$-$ & $-$ & $-$\tabularnewline
\bottomrule
\end{tabular*}
\end{table}

With consideration of SOC, the parity productions $\delta_{i}$ for the occupied states of $\mathrm{Ba}_{3}\mathrm{Cd}_{2}\mathrm{As}_{4}$
without counting their Kramer degenerate bands at the eight TRIM points are tabulated in Table \ref{tab:3}.
It is easy to find that $\mathrm{Ba}_{3}\mathrm{Cd}_{2}\mathrm{As}_{4}$ is a weak TI with
$Z_{2}$ invariants as $(0;110)$ \cite{fu2007topological}.

\begin{figure*}
\includegraphics[clip,scale=1, angle=0]{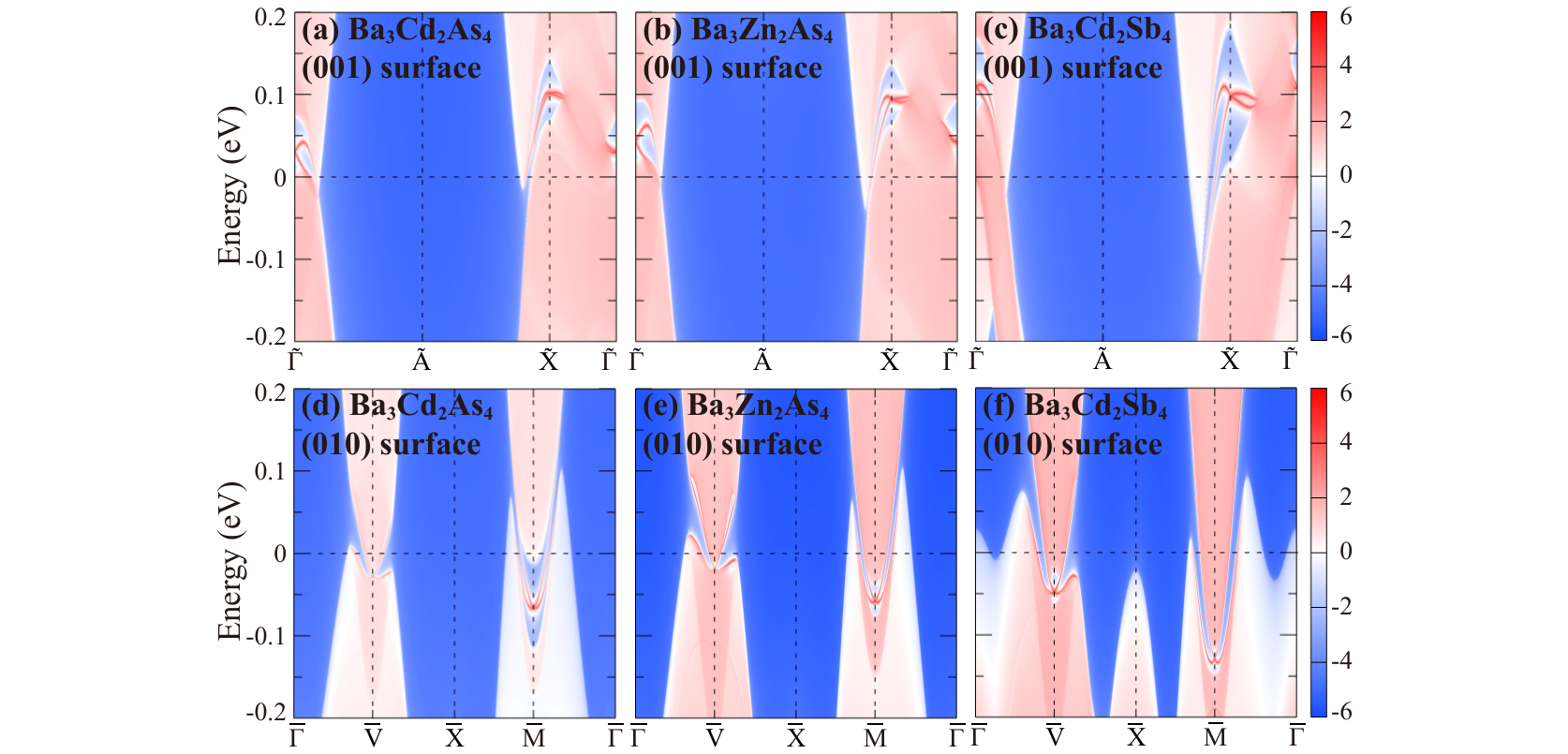}
\caption{(Color online) The projected surface states along the $\bold{k}_{c}$ and $\bold{k}_{b}$ directions of $\mathrm{Ba}_{3}\mathrm{Cd}_{2}\mathrm{As}_{4}$ ((a) and (d)), $\mathrm{Ba}_{3}\mathrm{Zn}_{2}\mathrm{As}_{4}$ ((b) and (e)) and $\mathrm{Ba}_{3}\mathrm{Cd}_{2}\mathrm{Sb}_{4}$ ((c) and (f)) with SOC obtained from the GGA calculations.
\label{spfig:3}}
\end{figure*}

The time-reversal polarization $\pi_{\tilde{k}}$ associated with surface momentum $\tilde{k}$ can be also calculated by these $\delta_{i}$ \cite{fu2007topological}. On the $(001)$ surface   (see Fig. \ref{spfig:4}(a)), the time-reversal polarization $\pi_{\tilde{\Gamma}}=\pi_{\tilde{X}}=0$ and $\pi_{\tilde{A}}=\pi$ indicate that there are two nontrivial surface Dirac points located at $\tilde{\Gamma}$ and $\tilde{X}$
points, respectively, as shown in Fig. \ref{spfig:3}(a).
On the $(010)$ surface (see Fig. \ref{spfig:4}(a)), the time-reversal polarization $\pi_{\bar{V}}=\pi_{\bar{M}}=0$ and $\pi_{\bar{\Gamma}}=\pi_{\bar{X}}=\pi$ indicate that there are two nontrivial surface Dirac points located at $\bar{\Gamma}$ and $\bar{X}$ points, respectively, as shown in Fig. \ref{spfig:3}(d).
The similar surface states are also got for $\mathrm{Ba}_{3}\mathrm{Zn}_{2}\mathrm{As}_{4}$ (see Figs. \ref{spfig:3}(b) and (e)) and $\mathrm{Ba}_{3}\mathrm{Cd}_{2}\mathrm{Sb}_{4}$ (see Figs. \ref{spfig:3}(c) and (f)).

\subsection{The Wilson loop flow of $\mathrm{Ba}_{3}\mathrm{Cd}_{2}\mathrm{As}_{4}$  and surface states of $\mathrm{Ba}_{3}\mathrm{Zn}_{2}\mathrm{As}_{4}$ and $\mathrm{Ba}_{3}\mathrm{Cd}_{2}\mathrm{Sb}_{4}$ on (1$\bar{1}$0) surface}

\begin{figure*}
\includegraphics{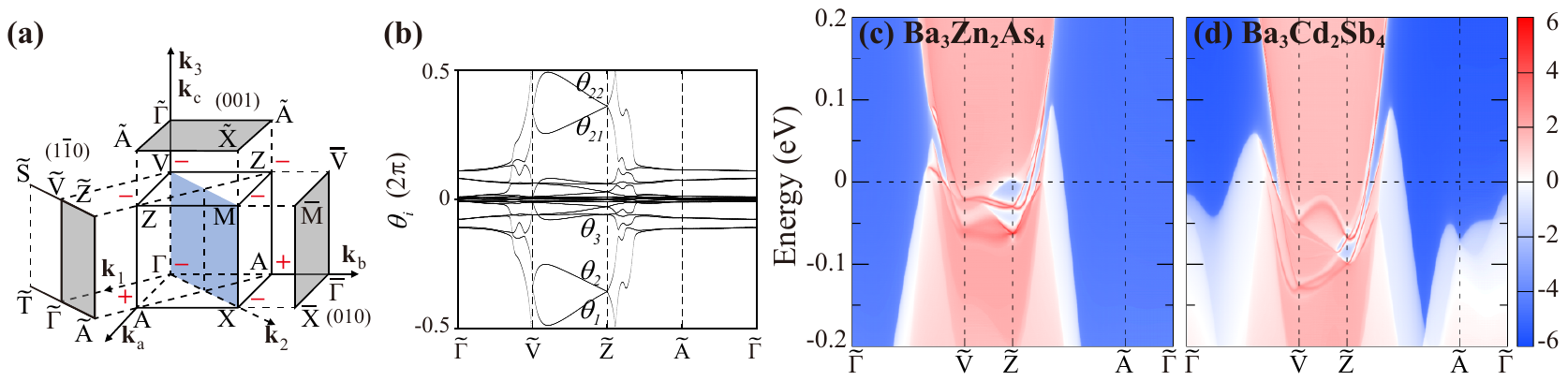}
\caption{(Color online) (a) The schematic diagram of the first BZ and the projected BZs. The $(1\bar{1}0)$, $(001)$ and $(010)$ surface BZs are projected along
 the $\bold{k}_1$, $\bold{k}_c$ and $\bold{k}_b$ directions, respectively.  The $+$ $(-)$ indicates the parity productions $\delta_{i}$ at eight TRIM points. (b) The Wilson loop flow of $\mathrm{Ba}_{3}\mathrm{Cd}_{2}\mathrm{As}_{4}$ along the high symmetrical paths in the (1$\bar{1}$0) surface BZ. The projected surface states along the $\bold{k}_{1}$  direction of (c) $\mathrm{Ba}_{3}\mathrm{Zn}_{2}\mathrm{As}_{4}$ and (d) $\mathrm{Ba}_{3}\mathrm{Cd}_{2}\mathrm{Sb}_{4}$ with SOC obtained from the GGA calculations.
\label{spfig:4}}
\end{figure*}

The Wilson loop flow of $\mathrm{Ba}_{3}\mathrm{Cd}_{2}\mathrm{As}_{4}$ along the high symmetrical paths in the (1$\bar{1}$0) surface BZ are shown in Fig. \ref{spfig:4}(b).
Due to time-reversal symmetry, $\theta_{i}(k_{\lambda})$ are double degenerate at TRIM of surface BZ and one can index them in pairs. Each pair contains one odd index $(i=1, 3, 5 \ldots {n_{occ}-1})$ and an even index $i+1$ when indexed in the ascending order of Berry phase. The two partners in one pair are separated along the high symmetrical lines other than at TRIM.
However, the two Wilson loop flow of $\theta_{i}(k_{\lambda})$ $(i=2, 4, 6 \ldots n_{occ})$ in one pair and $\theta_{i+1}(k_{\lambda})$ (when $i$=$n_{occ}$, $i$+1 takes 1) in the neighboring pair are double degenerate at two arbitrary momenta in the surface BZ. This is due to $C_{2}$ anomaly and protected by coexistence of $C_2$ rotation and time-reversal symmetry. For example, $\theta_{2}$ crosses $\theta_{3}$ at two generic momenta related by $C_2$ rotation. $\theta_{22}$ and $\theta_{1}$ also cross each other at another two generic momenta.

The $(1\bar{1}0)$ surface (see Fig. \ref{spfig:4}(a)) states of $\mathrm{Ba}_{3}\mathrm{Zn}_{2}\mathrm{As}_{4}$ and $\mathrm{Ba}_{3}\mathrm{Cd}_{2}\mathrm{Sb}_{4}$ are calculated with the consideration of SOC effect, as shown in Figs. \ref{spfig:4}(c) and (d), respectively. Because of the existence of the electron and hole pockets, the bulk gap is not clear around the $\widetilde{V}$ point. Only the trivial surface Dirac cone at $\widetilde{Z}$ can be clearly seen.

\subsection{The prism geometry and hinge states of $\mathrm{Ba}_{3}\mathrm{Cd}_{2}\mathrm{As}_{4}$}

\begin{figure*}
\includegraphics[clip,scale=1, angle=0]{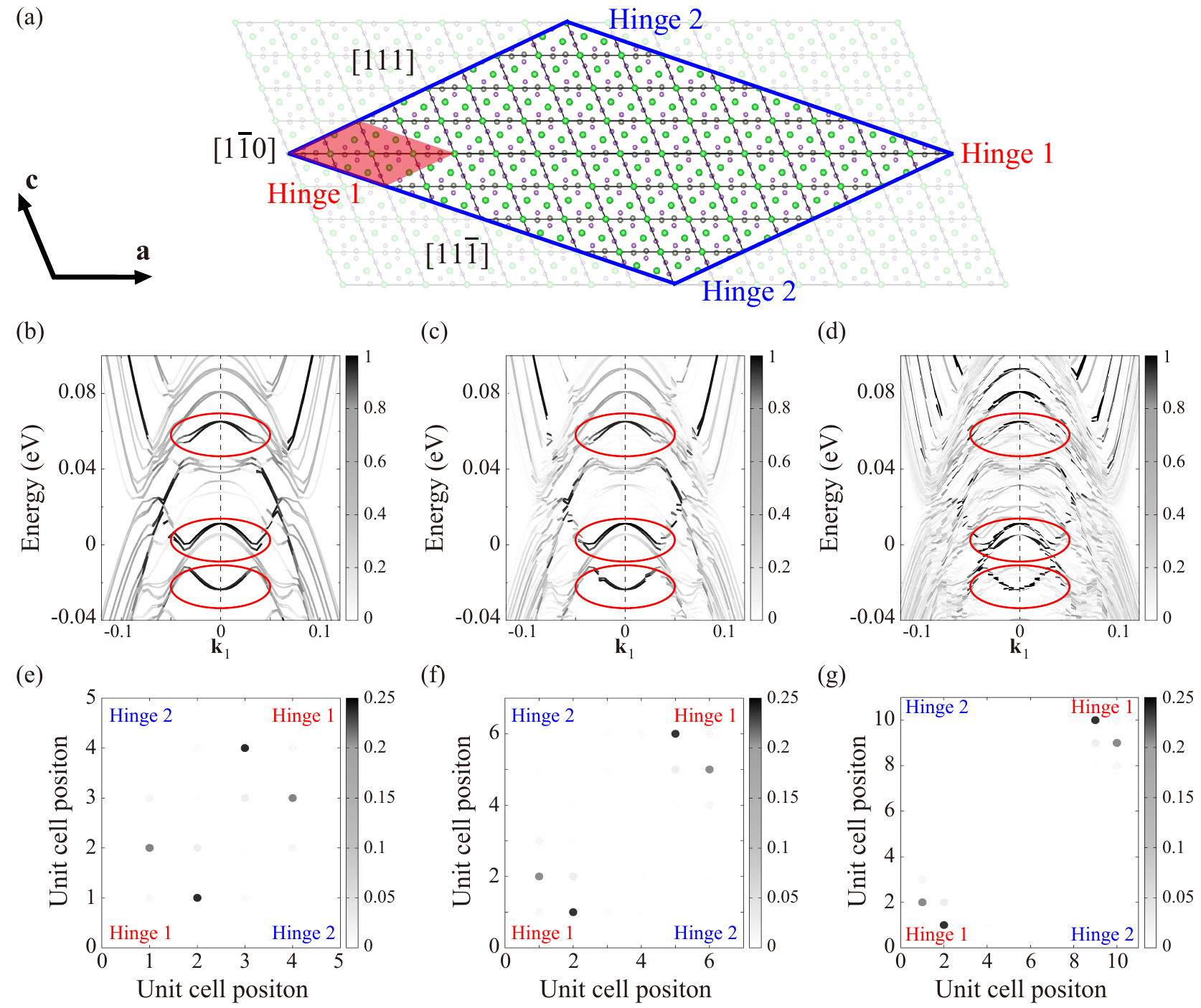}
\caption{(Color online) (a) The prism geometry with open boundary conditions around all four side surfaces in $[111]$ and $[11\bar{1}]$ directions, and periodic boundary condition along $[1\bar{1}0]$ direction in units of three primitive lattice vectors ($\bold{a}$, $\bold{b}$ and $\bold{c}$). The new unit cell is marked in red, and the supercell size of two open boundaries are 4\texttimes{}4 unit cells. (b), (c) and (d) The band structure of 4\texttimes{}4, 6\texttimes{}6 and 10\texttimes{}10 supercells with the color indicating the square of the projection of wave functions on the two hinges where the hinge states exist. The black bands marked by red circle represent helical hinge states, and the grey bands represent bulk and surface states. (e), (f) and (g) The square of the projection of hinge states wave functions on the unit cell of three supercells.
\label{spfig:5} }
\end{figure*}

The first prism geometry of $\mathrm{Ba}_{3}\mathrm{Cd}_{2}\mathrm{As}_{4}$ is shown in Fig. \ref{spfig:5}(a).
We construct a tight-binding model of the new unit cell and calculate the band structure of 4\texttimes{}4, 6\texttimes{}6 and 10\texttimes{}10 supercells, as shown in Fig. \ref{spfig:5}(b), (c) and (d). The hinge states marked by black lines appear as a Dirac cone protected by $C_{2}$ and time-reversal symmetry at TRIM. We find that the number of hinge sates don't change, while the numbers of surface states and bulk states increase greatly as the increasing of supercell size. Combining the results of Fig. \ref{spfig:5}(e), (f) and (g), we can see the distribution of hinge states in the supercell, and the hinge states are localized at the two hinges.

\begin{figure*}
\includegraphics[clip,scale=1, angle=0]{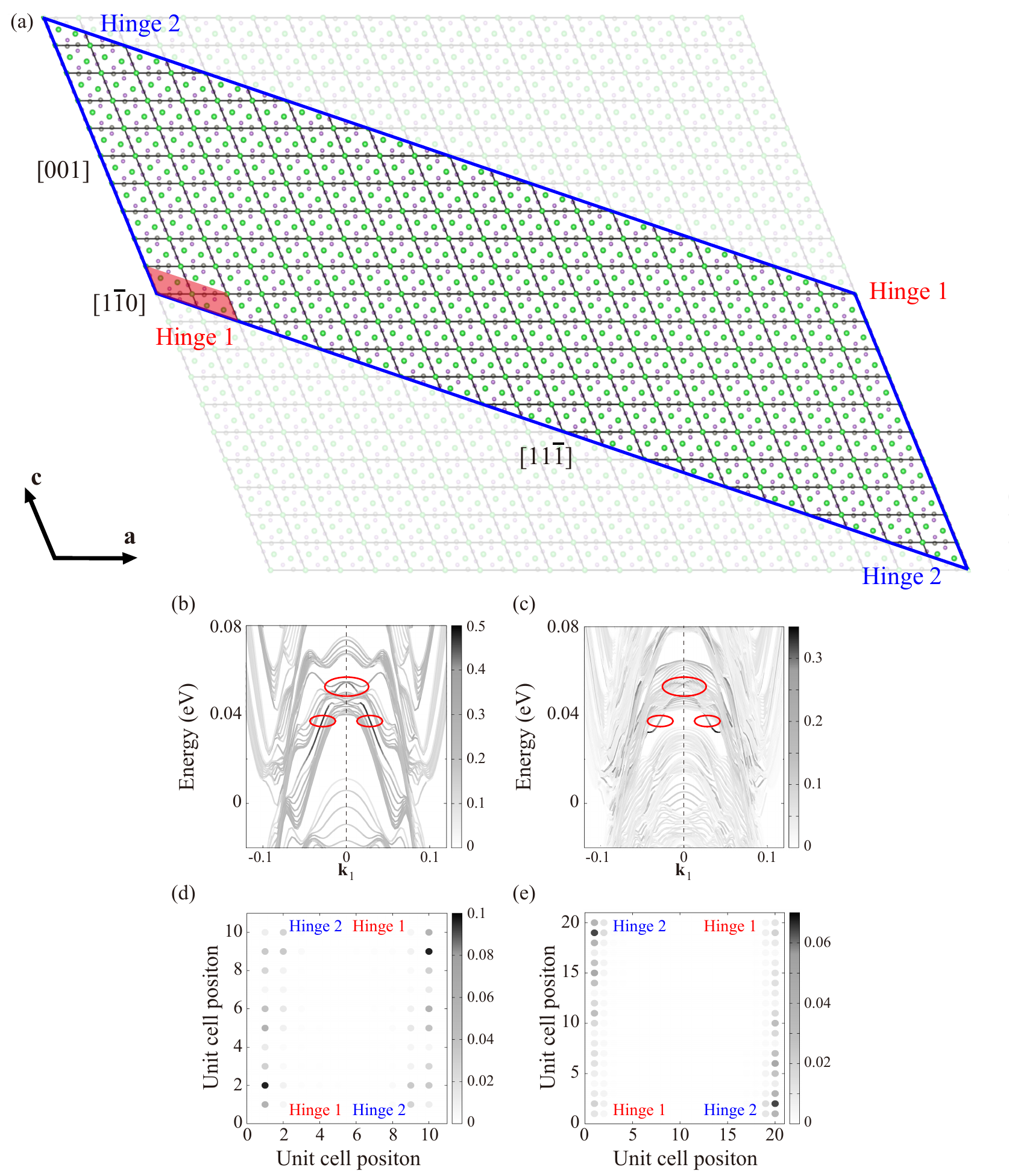}
\caption{(Color online) (a) The prism geometry with open boundary conditions around all four side surfaces in $[001]$ and $[11\bar{1}]$ directions, and periodic boundary condition along $[1\bar{1}0]$ direction in units of three primitive lattice vectors ($\bold{a}$, $\bold{b}$ and $\bold{c}$). The new unit cell is marked in red and the supercell size of two open boundaries are 10\texttimes{}10 unit cells. (b) and (c) The band structure of 10\texttimes{}10 and 20\texttimes{}20 supercells with the color indicating the square of the projection of wave functions on the two hinges where the hinge states exist. The black bands marked by red circle represent helical hinge states, and the grey bands represent bulk and surface states. (d) and (e) The square of the projection of hinge states wave functions on the unit cell of three supercells.
\label{spfig:6} }
\end{figure*}

To check the hinge states, we construct a different prism geometry and calculate the hinge states with 10\texttimes{}10 and 20\texttimes{}20 supercells, as shown in Fig. \ref{spfig:6}. We get the similar results. Interesting, we can see the distribution of hinge states are transferred from two hinges to the other two hinges when the size of supercell is enlarged from 10\texttimes{}10 to 20\texttimes{}20. The transfer of hinge states can occur, because there is no constrain on the position of hinge states by space symmetry and time-reversal symmetry.

\newpage

\bibliography{paperrefSM}